# Field-theory spin and momentum in water waves

K. Y. Bliokh[1], H. Punzmann[2], H. Xia[2], F. Nori[1,3], and M. Shats[2]

[1]*Theoretical Quantum Physics Laboratory, RIKEN Cluster for Pioneering Research, Wako-shi, Saitama 351-0198, Japan*
[2]*Research School of Physics, The Australian National University, Canberra, ACT 2601, Australia*
[3]*Physics Department, University of Michigan, Ann Arbor, Michigan 48109-1040, USA*

Spin is a fundamental yet nontrivial intrinsic angular-momentum property of quantum particles or fields, which appears within relativistic field theory. The spin density in wave fields is described by the theoretical Belinfante-Rosenfeld construction based on the difference between the canonical and kinetic energy-momentum tensors. These quantities have an abstract mathematical character and are usually considered as non-observable *per se*. Here we demonstrate, both theoretically and experimentally, that the Belinfante-Rosenfeld construction naturally arises in purely classical gravity (water surface) waves. There, the canonical momentum is associated with the generalized Stokes-drift phenomenon, while the spin is generated by subwavelength circular motion of water particles in inhomogeneous wave fields. Thus, we reveal the canonical spin and momentum in water waves and directly observe these fundamental field-theory properties as microscopic mechanical properties of particles in a classical wave system. Our findings shed light onto the nature of spin and momentum in wave fields, demonstrate the universality of field-theory concepts, and offer a new platform for studies of previously hidden aspects of quantum-relativistic physics.

> *The next waves of interest, that are easily seen by everyone and which are usually used as an example of waves in elementary courses, are water waves. As we shall soon see, they are the worst possible example, because they are in no respects like sound and light; they have all the complications that waves can have.*
>
> Richard P. Feynman

## INTRODUCTION

Spin angular momentum was first introduced to physics empirically in 1925 by Uhlenbeck and Goudsmit [1,2]. This allowed them to explain peculiarities of the emission spectra of solids and electron interactions with magnetic fields by a quantum "self-rotation" of electrons. Later, this new property was derived rigorously within the Dirac equation providing the quantum relativistic theory of electrons [3] and using relativistic field theory approaches [4–7]. Nowadays, spin is essential for numerous quantum and condensed-matter systems [2], ranging from basic properties of elementary particles and chemical elements, via widely used memory and tomography devices, to the advanced fields of spintronics [8,9] and quantum computing [10–12].

As early as in 1909, Poynting described the intrinsic angular momentum of circularly-polarized light (i.e., an electromagnetic wave with rotating electric and magnetic field vectors) [13]. This property was later observed via optical torque on matter, and it was associated with the spin of photons (i.e., relativistic massless quanta of light) [3,14–17]. Thus, the spin angular momentum naturally appears in classical electromagnetic fields [16–19] where it plays an important role in optical manipulation, light-matter interactions, information transfer, etc. [16,17,20–22]. In 1973, Jones argued that intrinsic angular momentum, or spin, can be ascribed to classical nonrelativistic waves, such as acoustic and internal gravity waves in fluids [23]. There spin is associated with the mechanical angular momentum of the medium particles oscillating or moving along microscopic elliptical orbits in wave fields. This interpretation was essentially neglected (Ref. [23] was cited only 8 times for almost half-century) and, recently, it was shown again that inhomogeneous acoustic (sound-wave) fields possess a nonzero spin angular momentum density [24–28]. This time, the presence of acoustic spin was supported by the analogy between acoustic and electromagnetic waves, and was confirmed experimentally [25].

Theoretically, various kinds of quantum and classical waves can be described within the corresponding field theories [3,6]. There, one of the main objects characterizing dynamical properties of the field is the energy-momentum tensor, which includes the field energy and momentum densities and provides for the local energy-momentum conservation laws. In 1940, Belinfante and Rosenfeld found a fundamental structure in this tensor, which explains the appearance of spin angular momentum and relates it to the momentum properties of the field [4–6]. They showed that there are canonical (non-symmetric, derived from the Noether theorem) and kinetic (symmetrized) versions of the energy-momentum tensor, which contain the corresponding canonical and kinetic momentum densities, $\mathbf{P}$ and $\mathbf{\Pi}$, related as

$$\mathbf{\Pi} = \mathbf{P} + \frac{1}{2}\nabla \times \mathbf{S}, \tag{1}$$

where $\mathbf{S}$ is the spin angular momentum density. This fundamental relation describes the appearance of spin in both quantum particles and classical (electromagnetic and acoustic) wave fields [6,7,17,19,21,23,25,28–30]. According to Noether's theorem, the integral values of both kinetic and canonical momentum densities are conserved in translation-invariant systems. In turn, rotational invariance is associated with the conservation of the integral angular momentum. Its density is given by $\mathbf{r} \times \mathbf{\Pi}$ in the kinetic picture and $\mathbf{r} \times \mathbf{P} + \mathbf{S}$ in the canonical one [6,19,28].

Despite such progress and thorough exploration of spin in various fields, this fundamental physical entity remains nontrivial and is described by rather abstract quantum-mechanical and relativistic-field-theory concepts [1–7]. Indeed, the "self-rotation" of the electron described by the Dirac spinors is far from an intuitively clear picture. Furthermore, the canonical momentum and spin densities in the field-theory relation (1) are usually regarded as unobservable *per se* [4–6], and only their integral values matter. In classical fields, rotating angular momentum properties underlying spin are more obvious, but rotating electric and magnetic fields in circularly-polarized light [13–22] or rotating medium particles in inhomogeneous sound waves [23–28] are never observed directly.

The purpose of this work is multifold. First, we will describe and observe the presence of spin in another kind of wave field, namely, in *gravity water-surface waves* [31]. We will show that the water-wave spin is described precisely by the same simple field-theory relation (1) involving the canonical and kinetic momenta. This is surprising because water-surface waves cannot be described by a relativistic Lagrangian field theory like electromagnetic or sound waves. This can be seen from the fact that electromagnetic and acoustic field theories are based on the properties of the Minkowski spacetime and essentially involve linear dispersion $\omega = ck$ ($c$ is the speed of light or sound), while water waves are inherently dispersive: e.g., $\omega = \sqrt{gk}$ in the deep-water approximation ($g$ is the gravitational acceleration). Although there is a number of rather sophisticated Lagrangian and Hamiltonian approaches to fluid dynamics and water waves [32–35], they do not provide a simple unified picture of momentum and angular momentum of surface gravity waves, and, in contrast to their electromagnetic and acoustic counterparts, these fundamental quantities are almost never mentioned in textbooks on fluid dynamics (Supplementary Materials) and do not typically appear in experimental observations. Here we argue that concepts of spin and kinetic/canonical momenta, originating from relativistic field theory, illuminate and accurately describe the observable dynamical properties of surface gravity waves.

Second, we will provide the direct observation of the motion of water particles underlying the spin and canonical-momentum densities (1). In doing so, the rotational motion of particles corresponds to the spin density $\mathbf{S}$, whereas the translational motion due to the *generalized Stokes drift* [36–38] corresponds to the canonical momentum density $\mathbf{P}$. To the best of our knowledge, this is the first direct observation of the *microscopic* origin of the spin angular momentum and canonical momentum in wave fields. Importantly, the generalized Stokes drift described and observed in our work accurately characterizes the mass transport in acoustic and water wavefields and provides the directly observable momentum of these waves. This is crucial for numerous applications involving transport of microscopic and macroscopic objects in water waves.

Finally, by comparing our approach to water waves with other wave theories, we demonstrate the universality of the spin, momentum, and Belinfante-Rosenfeld concepts across quantum systems, electromagnetism, acoustics, and hydrodynamics (even though relativistic field theory is not directly applicable to water waves). This opens up new opportunities for both quantum-relativistic and classical physics.

## RESULTS

### Basic spin and momentum properties of vector wave fields

To begin with, Table I lists the main dynamical quantities involved in Eq. (1), as well as the energy density, for monochromatic electromagnetic waves in free space [17–19,21,28–30] and sound waves in a fluid or gas [25–28]. For all kinds of monochromatic waves, we consider complex space-dependent field amplitudes $\mathbf{F}(\mathbf{r})$, so that real time-dependent fields are $\mathsf{F}(\mathbf{r},t) = \mathrm{Re}\left[\mathbf{F}(\mathbf{r})e^{-i\omega t}\right]$, where $\omega$ is the wave frequency. In this manner, electromagnetic waves are described by the complex electric and magnetic fields, $\mathbf{E}(\mathbf{r})$ and $\mathbf{H}(\mathbf{r})$, while acoustic waves are described by the complex velocity field $\mathbf{v}(\mathbf{r})$ and scalar pressure field $p(\mathbf{r})$. In both electromagnetic and acoustic cases, the canonical momentum density $\mathbf{P}$ is determined by the form $\mathrm{Im}\left[\mathbf{F}^* \cdot (\nabla)\mathbf{F}\right]$ entirely similar to the probability current in quantum mechanics, i.e., the local "expectation value" of the canonical quantum-mechanical momentum operator $-i\nabla$ [29,30]. In turn, the spin angular momentum density $\mathbf{S}$ is determined by the form $\mathrm{Im}(\mathbf{F}^* \times \mathbf{F})$ which points into the direction normal to the polarization ellipse of the field $\mathbf{F}$ and is proportional to its ellipticity [16,17,39].

Notably, both electromagnetic and acoustic canonical momentum and spin densities in monochromatic fields are measurable via radiation forces and torques on small absorbing particles [17,25,27,30]. Note also that spatial integrals of the spin densities for localized circularly-polarized paraxial electromagnetic waves and sound wavefields are in agreement with the quantum-mechanical spin values of $\hbar$ per photon [16,17] and 0 per phonon [26,28]. Substituting the canonical momentum and spin densities into Eq. (1) and using the equations of motion for the wave fields (i.e., Maxwell and acoustic wave equations), one obtains the kinetic momentum density $\mathbf{\Pi}$. It is given by the well-known Poynting vector for electromagnetic waves and its acoustic analogue for sound waves [31].

Importantly, the acoustic spin and canonical momentum densities can be immediately associated with the mechanical properties of microscopic particles of the medium. Generally, such particles experience a combination of rotational and translational motion in the sound-wave field. First, the microscopic periodic motion of the medium particles is generically *elliptical* and corresponds to the polarization of the vector velocity field $\mathbf{v}$. The oscillating velocity field $\mathbf{v}e^{-i\omega t}$ corresponds to the displacement field $\mathbf{a}e^{-i\omega t} = i\omega^{-1}\mathbf{v}e^{-i\omega t}$, which yields the time-averaged mechanical angular momentum density $(\rho/2)\text{Re}(\mathbf{a}^* \times \mathbf{v})$ (where $\rho$ is the mass density of the medium) [23,25,28], precisely equivalent to the spin density $\mathbf{S}$ in Table I. Second, the medium particles in a sound-wave field can experience the slow *Stokes drift* [36–38], a phenomenon known in hydrodynamics for surface water waves and related to the difference between the Eulerian and Lagrangian velocities of the particles. (A somewhat related phenomenon of the transformation of an oscillatory motion to a linear drift is known as acoustic streaming [40], with numerous examples in acoustofluidics and surface acoustic waves [41,42].) So far, the Stokes drift was described only for plane water surface waves with vertical inhomogeneity, while here we generalize this phenomenon to arbitrary inhomogeneous monochromatic fields. The momentum density associated with the generalized Stokes drift can be written as (Supplementary Materials) $(\rho/2)\text{Re}\left[(\mathbf{a}^* \cdot \nabla)\mathbf{v}\right]$, which for sound waves with $\nabla \times \mathbf{v} = \mathbf{0}$ yields the *canonical momentum density* in Table I. This expression is similar to the '*pseudomomentum*' of waves in fluids or gases introduced in 1978 by Andrews and McIntyre [43]. Note, however, that the oscillatory and drift motions of particles in bulk sound waves are difficult to observe directly due to the very small displacements $\mathbf{a}$ in typical sound wave fields.

| | **Electromagnetism** | **Acoustics** | **Water waves** |
|---|---|---|---|
| **Wave fields** | electric $\mathbf{E}$, magnetic $\mathbf{H}$ | velocity $\mathbf{v}$, pressure $p$ | In-plane velocity $\mathbf{V}$, vertical velocity $W$ |
| **Energy density** $U$ | $\frac{1}{4}\left(\varepsilon|\mathbf{E}|^2 + \mu|\mathbf{H}|^2\right)$ | $\frac{1}{4}\left(\beta|p|^2 + \rho|\mathbf{v}|^2\right)$ | $\frac{\rho}{4}\left(3|W|^2 + |\mathbf{V}|^2\right)$ |
| **Kinetic momentum density** $\mathbf{\Pi}$ | $\frac{1}{2c^2}\text{Re}(\mathbf{E}^* \times \mathbf{H})$ | $\frac{1}{2c_s^2}\text{Re}(p^*\mathbf{v})$ | $\frac{\rho k}{\omega}\text{Im}(W^*\mathbf{V})$ |
| **Canonical momentum density** $\mathbf{P}$ | $\frac{1}{4\omega}\text{Im}\left[\varepsilon\mathbf{E}^* \cdot (\nabla)\mathbf{E} + \mu\mathbf{H}^* \cdot (\nabla)\mathbf{H}\right]$ | $\frac{\rho}{2\omega}\text{Im}\left[\mathbf{v}^* \cdot (\nabla)\mathbf{v}\right]$ | $\frac{\rho}{2\omega}\text{Im}\left[\mathbf{V}^* \cdot (\nabla_2)\mathbf{V} + W^*\nabla_2 W\right]$ |
| **Spin AM density** $\mathbf{S}$ | $\frac{1}{4\omega}\text{Im}(\varepsilon\mathbf{E}^* \times \mathbf{E} + \mu\mathbf{H}^* \times \mathbf{H})$ | $\frac{\rho}{2\omega}\text{Im}(\mathbf{v}^* \times \mathbf{v})$ | $\frac{\rho}{2\omega}\text{Im}(\mathbf{V}^* \times \mathbf{V})$ |

**Table I. The energy, momentum, and spin properties of electromagnetic, acoustic, and deep-water gravity monochromatic wavefields.** Here $c = 1/\sqrt{\varepsilon\mu}$ is the speed of light, $c_s = 1/\sqrt{\beta\rho}$ is the speed of sound, $\varepsilon$ and $\mu$ are the permittivity and permeability of the electromagnetic medium, and $\rho$ and $\beta$ are the mass density and compressibility of the acoustic medium or fluid.

## Spin and momentum of gravity water waves

We now consider a wave system which is not typically associated with relativistic field theories and spin: water-surface (gravity) waves [31]. Deep-water gravity waves are characterized by the dispersion $\omega^2 = kg$ ($g$ is the gravitational acceleration, $k$ is the wave number), and all wave fields decay exponentially from the unperturbed water surface $z = 0$ deep into the water $z<0$ as $\propto \exp(kz)$ [31]. Thus, in contrast to the 3D electromagnetism and acoustics, this system is quasi-2D. Since unperturbed water is translationally symmetric in the $(x,y)$ plane, and rotationally symmetric about the $z$-axis, it is natural to expect a conserved 2D momentum and $z$-directed angular momentum of water waves. Therefore, we separate the 3D velocity $\mathbf{v}$ of the water particles in the gravity-wave field into the in-plane 2D vector $\mathbf{V}=(v_x,v_y)$ and the normal component $W=v_z$. We will focus on the 2D motion of surface water particles in the $(x,y)$ plane ($z=0$), but will also take into account all physical properties related to the vertical $z$-motion. The 2D gradient (momentum) operator is $\nabla_2 =(\partial_x,\partial_y)$, while the vector product (spin) operator "$\times$" acting in the plane can only produce a $z$-directed vertical spin.

Since the motion of water particles in the oscillating 2D velocity field $\mathbf{V}e^{-i\omega t}$ is entirely similar to the motion of medium particles in the oscillating sound-wave field $\mathbf{v}e^{-i\omega t}$, the $z$-directed angular momentum density can be written akin to the acoustic spin density:

$$\mathbf{S}=\frac{\rho}{2\omega}\mathrm{Im}\left(\mathbf{V}^*\times\mathbf{V}\right). \qquad (2)$$

This provides the *spin density for gravity waves* associated with polarization of the vector field $\mathbf{V}$, see Table I. This spin appears in inhomogeneous (e.g., interference) water-wave fields, because of the circular (or, generically, elliptical) motion of water particles in the $(x,y)$ plane. Note that the spin (2) considered in our work is *not* the spin angular momentum considered by Longuet-Higgins in Ref. [44]. The latter one is related to the elliptical motion of water particles in the propagation $(z,x)$ plane (for $x$-propagating plane waves) and is directed along the horizontal $y$-axis; our vertical spin (2) vanishes in a plane wave.

Next, the water particles experience the Stokes drift [36–38]. So far, this phenomenon has been known for the circular motion of water particles in the plane orthogonal to the water surface, i.e., involving the vertical velocity component $W$. For inhomogeneous wave fields with a nonzero spin $\mathbf{S}$, the particles can also exhibit elliptical orbits in the projection onto the water-surface plane. This produces the Stokes drift described by the in-plane velocity $\mathbf{V}$. Calculating the total Stokes drift in an arbitrary monochromatic gravity-wave field, we obtain that its velocity $\mathbf{u}$ is given by (see Supplementary Materials):

$$\mathbf{u}=\frac{1}{2\omega}\mathrm{Im}\left[\mathbf{V}^*\cdot\left(\nabla_2\right)\mathbf{V}+W^*\nabla_2 W\right], \qquad \mathbf{P}=\rho\mathbf{u}. \qquad (3)$$

Here, multiplying this Stokes drift velocity by the mass density, we obtained the *canonical momentum density* $\mathbf{P}$ *for gravity waves*, analogous to the 'pseudomomentum' by Andrews and McIntyre [43], see Table I. Importantly, the Stokes drift, i.e., the canonical momentum, produces *mass transport* in water waves [37], such as, e.g., the driftwood along the ocean coasts [45]. For plane waves, , $\mathbf{v}\propto\exp(i\mathbf{k}\cdot\mathbf{r})$, the generalized Stokes drift (3) is proportional to the wavevector: $\mathbf{u}\propto\mathbf{k}$. This provides the natural similarity between the canonical momentum (3) and de Broglie momentum in quantum mechanics.

Now, substituting the above canonical momentum and spin densities into the Belinfante-Rosenfeld relation (1), we obtain the kinetic momentum density $\mathbf{\Pi}=(\rho k/\omega)\mathrm{Im}(W^*\mathbf{V})$ for surface water waves, see Table I. Remarkably, its form is equivalent to the conserved water-wave momentum derived by Peskin [46] (Supplementary Materials). It should be noticed that the energy and momentum conservation laws for water waves are rather nontrivial, because they essentially involve $z$-integrals of generic time-dependent fields [46]. They are reduced to simple forms listed in Table I only for the case of monochromatic fields, when all fields decay as $\propto \exp(kz)$, and the $z$-integrals of quadratic forms are evaluated as $\int_{-\infty}^{0} ... dz = (2k)^{-1}...$.

Moreover, unlike electromagnetic and acoustic waves, water-surface waves *cannot* be described within a relativistic Lagrangian field theory. This can be seen from the fact that these waves are essentially dispersive, $\omega = \sqrt{gk}$, which breaks the Minkowski spacetime structure $\omega = ck$ underlying electromagnetic and acoustic field theories. It is not by chance that the electromagnetic field Lagrangian and conservation laws are described in any textbook in electromagnetism; the simplest energy conservation for sound waves can also be found in textbooks [31] although the acoustic field Lagrangian and other conservation laws are only present in more specialized literature; whereas water-surface wave Lagrangian and conservation laws are absent in textbooks in hydrodynamics. (Also, one of the most important works on water-wave momentum, Ref. [43], only shows the energy conservation law for bulk acoustic waves.) We summarize the main differences between electromagnetic, sound, and water-surface waves in the Supplementary Materials. In agreement with the epigraph to this work, water waves are the worst possible system for application of a relativistic field theory. Probably, the easiest ways to derive dynamical properties of water-surface waves are: (i) to use macroscopic wave equations of motion (as done by Peskin [46] in deriving the conserved kinetic momentum $\mathbf{\Pi}$) or (ii) to involve microscopic mechanical properties of water particles moving in wave fields (as done here in deriving canonical momentum $\mathbf{P}$ and spin $\mathbf{S}$). Strikingly, both ways result in the kinetic and canonical densities exactly satisfying the Belinfante-Rosenfeld relation (1) lying at the heart of relativistic field theory. This suggests that the Belinfante-Rosenfeld construction has a deeper origin than a standard relativistic field theory [23].

Most importantly, the spin and momentum densities (2) and (3) are not abstract theoretical quantities but rather observable dynamical properties of surface gravity waves. We proceed with the direct experimental observation of these fundamental properties in structured water waves.

## Examples and experimental measurements

We are now in a position to show explicit examples of surface gravity waves with nonzero spin and momentum. The first example is a simple interference of two plane waves with equal frequencies and orthogonal wavevectors $\mathbf{k}_1 \perp \mathbf{k}_2$. The spin and momentum in two-wave interference has been previously considered for optical and sound waves [25,28,47]. Choosing the $y$-axis to be directed along $\mathbf{k}_1 + \mathbf{k}_2$, the spin and canonical-momentum densities, Eqs. (2) and (3), yield (Supplementary Materials):

$$\mathbf{S} \propto -\overline{\mathbf{z}} \sin \tilde{x}, \quad \mathbf{P} \propto \overline{\mathbf{y}}(2 + \cos \tilde{x}), \tag{4}$$

where $\tilde{x} = \sqrt{2} kx$ and the overbar indicates the unit vectors of the corresponding axes. The distributions of these densities, together with the numerically calculated microscopic water-particle trajectories, are shown in Fig. 1B. One can clearly see that the canonical momentum density corresponds to the Stokes drift of the particles (which everywhere occurs in the $y$-direction), whereas the spin density corresponds to the microscopic elliptical motion of particles (which has alternating $x$-dependent directions).

We have performed an experiment demonstrating the above motion of water particles and, thereby, the presence of canonical momentum and spin in the two-wave interference, Fig. 1. The experimental setup is shown in Fig. 1A (see Supplementary Materials for details). Interfering gravity waves were generated in a wave tank of size $1.0 \times 0.6$ m$^2$ and depth $h = 0.1$ m by two orthogonal paddles driven by two synchronized computer-controlled shakers. We worked with the wave frequencies $\omega / 2\pi \in (3,9)$ Hz which corresponds to the wavelengths $2\pi / k \in (0.03, 0.17)$ m satisfying the deep-water condition $\tanh(kh) \simeq 1$. Fluid motion at the water surface was visualized using buoyant tracer particles (Polyamid, 50 μm) illuminated by a LED panel placed underneath the transparent wave tank. A video camera on top was used to capture the motion of the tracer particles.

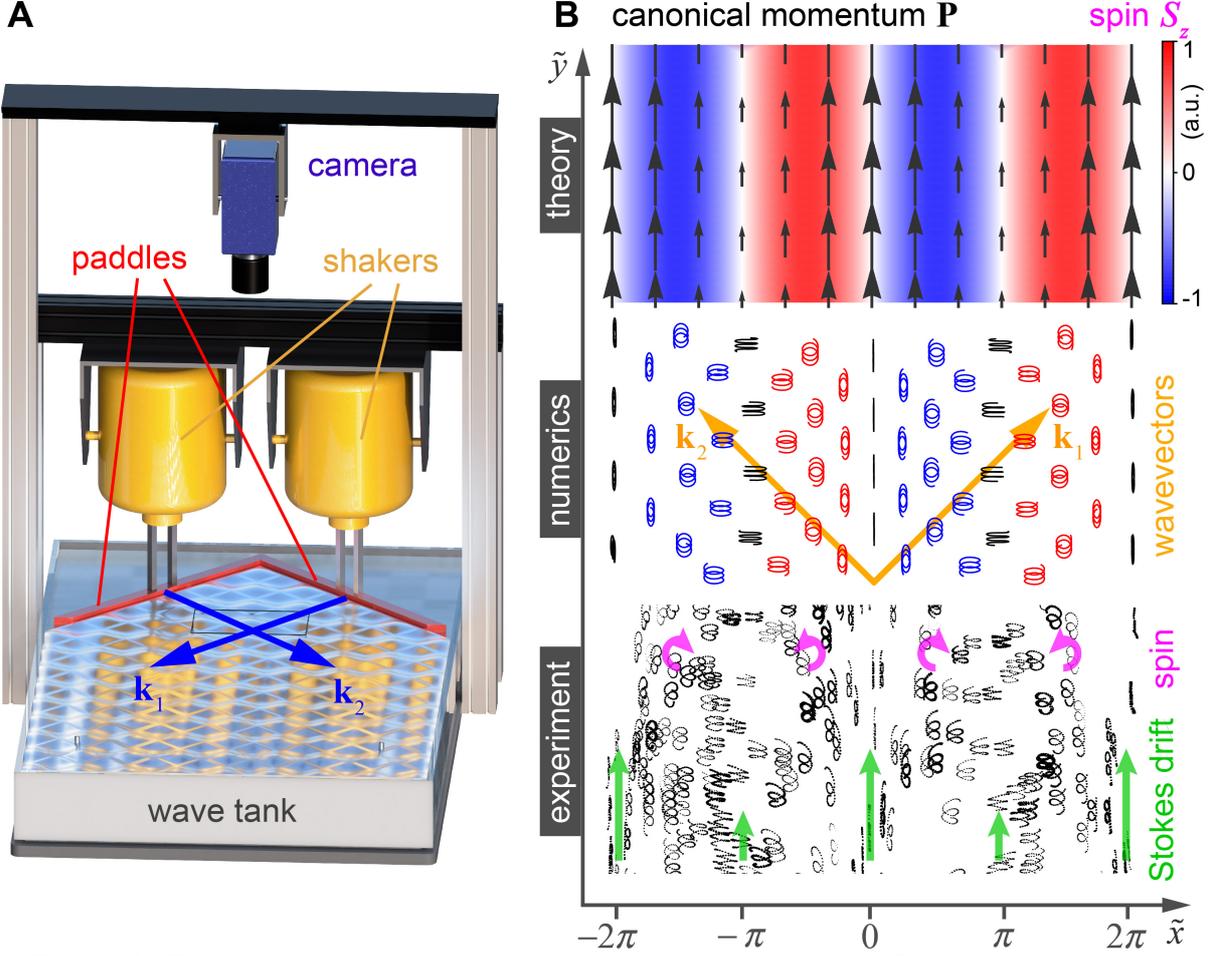

**Figure 1**. **The canonical momentum and spin densities in interfering surface water waves.** (**A**) Schematic of the experimental setup for the observation of the particle motion in interfering gravity waves. (**B**) Spin and momentum properties of two interfering gravity waves with equal frequencies, amplitudes, and orthogonal wavevectors $\mathbf{k}_1$ and $\mathbf{k}_2$. The theoretical plot shows the distributions of the canonical momentum density **P** and spin density **S**, Table I. Numerical and experimental plots depict trajectories of microscopic particles for three wave periods $6\pi/\omega$. The Stokes drift of the particles and their circular motion correspond to the canonical momentum and spin, respectively. Parameters are: $\tilde{x} = \sqrt{2}kx$, $\tilde{y} = \sqrt{2}ky$, and $\omega/2\pi = 6$ Hz.

In Fig. 1B, one can see that the experimentally measured trajectories are very similar to the numerically-calculated ones. To show that these experimental observations are in quantitative agreement with the theoretical spin and momentum densities, we measure the spatial and frequency dependences of the drift velocities and rotational radii of the particles. First, the canonical momentum density should behave as $P_y \propto k/\omega \propto \omega$ because the gradient operator scales as $\propto k$. Obviously, the particle drift velocity $u$ should obey the same frequency dependence. Second, the spin density is inversely proportional to the frequency: $S_z \propto \omega^{-1}$. As we have discussed, the spin can be associated with the mechanical angular momentum of water particles. At the points of maximum absolute value of the spin, $\tilde{x} = \pm\pi/2$, the water particles follow near-circular orbits of radius $a$, see Fig. 1B, and their angular momentum is $\propto a^2\omega$. Therefore, this radius should depend on the frequency as $a \propto \omega^{-1}$. Figure 2 shows the experimentally measured dependences $u(\omega)$ and $a(\omega)$ for water particles. These dependences are in excellent agreement with the above theoretical predictions and the $x$-dependence $\propto (2 + \cos \tilde{x})$ of the canonical momentum. The only discrepancy is that the drift velocity is offset by a constant value such that $u(0) \neq 0$. This is due to the presence of small return flows in the finite-size wave tank (Supplementary Materials).

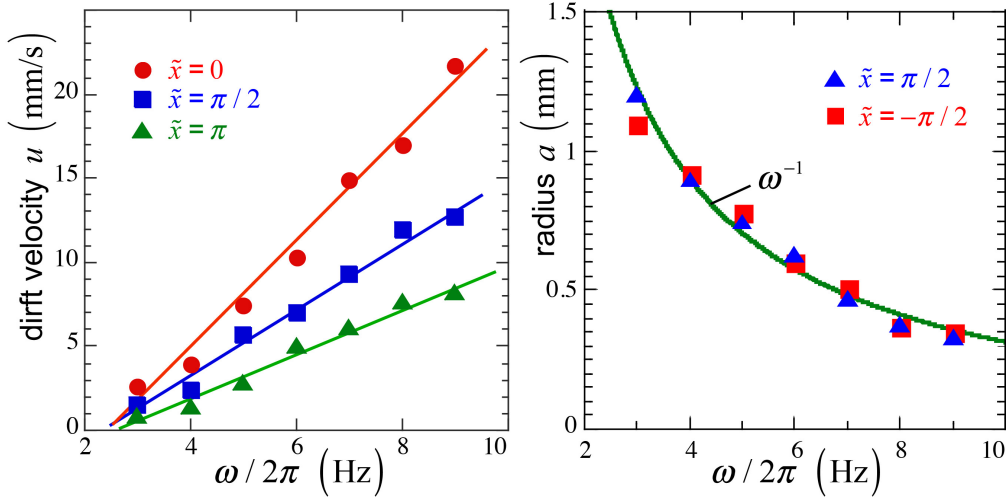

**Figure 2**. **Frequency dependencies of the Stokes drift and the microscopic orbits in interfering gravity waves from Fig. 1.** The experimentally measured Stokes drift velocity grows linearly with the wave frequency and depends on the position $\tilde{x} = \sqrt{2} kx$. The radii of the circular motion of water particles at the maxima of the spin density are inversely proportional to the wave frequency. These dependences are in agreement with theoretical predictions based on the canonical momentum and spin densities.

As another example, we consider an interference of two orthogonal standing water waves with equal amplitudes and frequencies, which is equivalent to four propagating waves. In this case, the spin density (2) forms a periodic chessboard-like structure, whereas the canonical momentum density (3) forms vortex-like flows around the maxima and minima of the spin density [48] (see Supplementary Materials and Fig. 3):

$$\mathbf{S} \propto \overline{\mathbf{z}} \sin\varphi \cos\tilde{x} \cos\tilde{y}, \qquad \mathbf{P} \propto \sin\varphi (\overline{\mathbf{y}} \sin\tilde{x} \cos\tilde{y} - \overline{\mathbf{x}} \cos\tilde{x} \sin\tilde{y}). \qquad (5)$$

Here, $\tilde{x} = kx$, $\tilde{y} = ky$, and $\varphi$ is the phase between the two orthogonal standing waves. Figure 3 shows the numerically calculated and experimentally measured trajectories of microscopic particles in the interference of two orthogonal standing waves with $\varphi = \pi/2$ (the spinless case $\varphi = 0$ is shown in Supplementary Materials). One can see that particles follow large wavelength-scale vortex-like orbits due to the Stokes drift associated with the momentum $\mathbf{P}$. Simultaneously, the particles experience microscopic elliptical motion around their current positions, which produces the local angular momentum associated with the spin $\mathbf{S}$. We emphasize that the two orbital motions here have different scales and qualitatively different nature. Indeed, the radius of the microscopic spin-related circular motion is proportional to the *amplitude* of the wave and can be made as small as needed, while the radius of the macroscopic vortex-like motion is fixed by the *wavelength*.

## DISCUSSION

To conclude, we have revealed the fundamental spin and momentum properties in water-surface (gravity) waves. Surprisingly, these quantities are precisely described by the relativistic field-theory construction by Belinfante-Rosenfeld [4–6], which underpins the spin and momentum of quantum and classical particles and fields [17,19,21,25,28–30]. We have shown that the canonical momentum density in acoustic and water waves can be directly associated with the mass transfer due to the generalized Stokes drift [36–38], while the spin density originates from the mechanical angular momentum of the medium particles following microscopic elliptical trajectories. Most importantly, we have provided the direct observation of these drift and rotational dynamics of water particles in inhomogeneous gravity-wave fields. This can be regarded as the first direct observation of the microscopic origin of the canonical spin and momentum in structured wavefields.

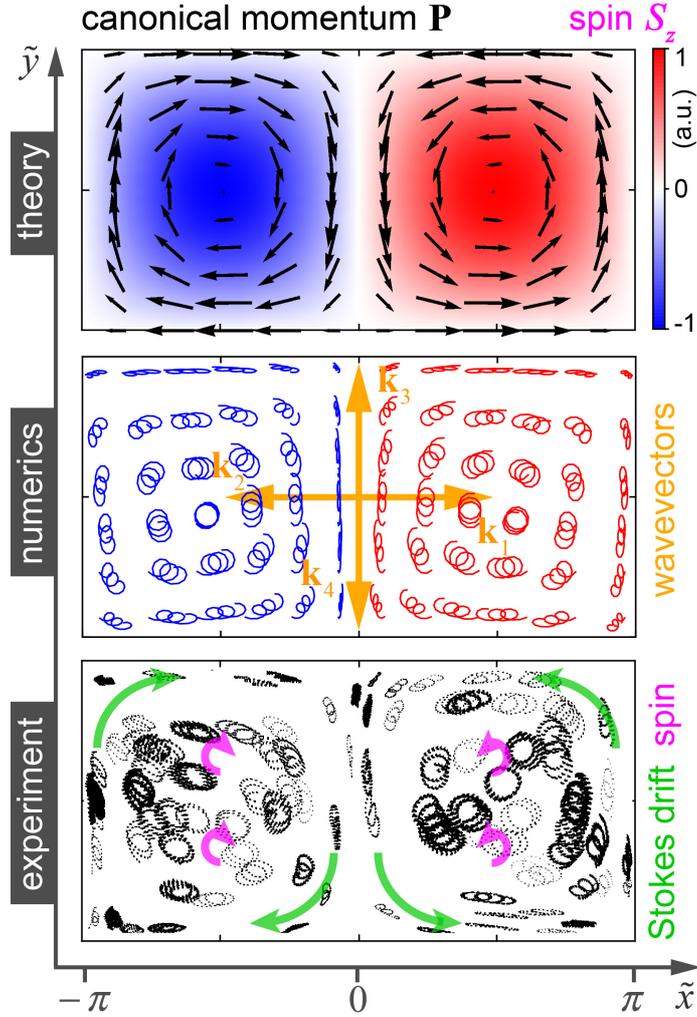

**Figure 3. Canonical momentum and spin densities in the interference of standing gravity waves.** Same as in Fig. 1 but for two interfering orthogonal standing waves with equal frequencies and amplitudes (i.e., equivalently, four propagating plane waves with the wavevectors $\mathbf{k}_{1,2,3,4}$). Parameters are: $\tilde{x} = kx$, $\tilde{y} = ky$, and $\omega/2\pi = 5.3$ Hz.

The appearance of a relativistic field-theory construction in the properties of water-surface waves is rather surprising. Such waves can be associated with relativistic field theory neither physically nor mathematically. Their dispersion is inconsistent with the Minkowski-like spacetime symmetries, whereas a simple (2+1)D form of the equations of motion does not possess even the basic energy conservation law (Supplementary Materials). Nonetheless, the presence of the (2+1)D space-time symmetries and microscopic mechanical description of the motion of the medium molecules in water-wave fields allows one to obtain meaningful $(x,y)$-momentum and $z$-directed angular momentum of water waves. Surprisingly, these quantities involve $z$-directed spin and are exactly described by the Belinfante-Rosenfeld relation. This hints that the Belinfante-Rosenfeld relation has a more fundamental origin than relativistic field theory.

Our results can have a multifold interdisciplinary impact. They shed light onto the nature of spin and momentum in various wave fields and illuminate the universality of field-theory relations, which so far have been considered as abstract theoretical quantities underlying observable physical phenomena on a higher level. Our findings also unveil the nontrivial nature of water-wave and acoustic momentum, which caused longstanding discussions [43,46,49,50]. The presence of nonzero spin density explains the existence of two (canonical and kinetic) momenta, as well as the direct observability of at least one of these. Our experiments provide direct measurements of the local mass transport in structured water-surface waves. Notably, using the dynamical spin and momentum concepts, one can produce structured water-wave fields for desired manipulation of particles, including transport and rotation, akin to optical manipulations [16,17,27,51]. Furthermore, the interplay of spin and linear momentum can produce novel types of transport in water waves. In particular, considering propagation of spinning particles in two interfering water waves, Fig. 1, we found that they exhibit a large-scale transverse splitting of opposite spins, as shown in Fig. 4. This is a water-wave analogue of the *spin-Hall effect*, a universal manifestation of *spin-orbit interactions* known in condensed-matter physics [52], optics [20], and even having implications in hydrodynamics [53]. Thus, our work offers a new platform for simulations and applications of various spin-related phenomena using readily accessible classical waves.

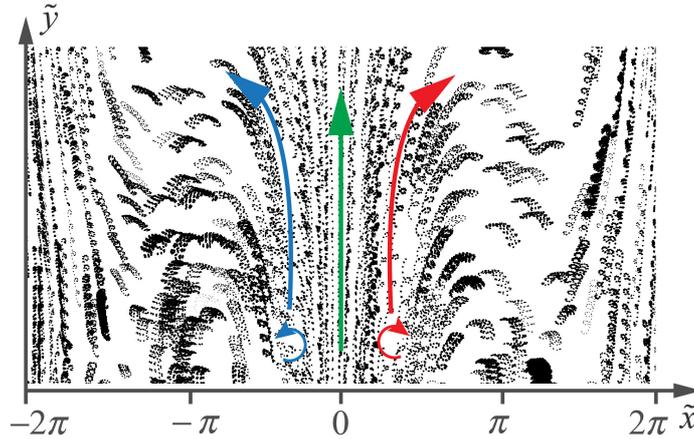

**Figure 4**. **Water-wave analogue of the spin-Hall effect.** Experimentally measured trajectories of water particles in two interfering waves, same as in Fig. 1B but for a longer period of time (about 10 wave periods). The forward propagation of spinning particles due to the Stokes drift is accompanied by the transverse splitting of oppositely spinning particles.

**Acknowledgements:** We acknowledge fruitful discussions with Lucas Burns and Justin Dressel. This work was partially supported by the Australian Research Council (via the Discovery Grant No. DP190100406 and Linkage Grant No. LP160100477), NTT Research, Army Research Office (ARO) (Grant No. W911NF-18-1-0358), Japan Science and Technology Agency (JST) (via the CREST Grant No. JPMJCR1676), Japan Society for the Promotion of Science (JSPS) (via the KAKENHI Grant No. JP20H00134 and the grant JSPS-RFBR Grant No. JPJSBP120194828), the Grant No. FQXi-IAF19-06 from the Foundational Questions Institute Fund (FQXi), and a donor advised fund of the Silicon Valley Community Foundation.